# Anisotropic Hot Carrier Relaxation and Coherent Phonon Dynamics in Type-II Weyl Semimetal TaIrTe$_4$

*Zheng Zhu[1], Jingwen Wang[1], Hao Yu[2], Jialin Lu[1], Tianshu Lai[1], Peng Yu[2, #], Tianran Jiang[1, #] and Ke Chen[1, #]*

[1]Center for Neutron Science and Technology, Guangdong Provincial Key Laboratory of Magnetoelectric Physics and Devices, State Key Laboratory of Optoelectronic Materials and Technologies, School of Physics, Sun Yat-sen University, 510275, Guangzhou, China

[2]State Key Laboratory of Optoelectronic Materials and Technologies, School of Materials Science and Engineering, Sun Yat-sen University, 510275, Guangzhou, China

# Corresponding authors: chenk35@mail.sysu.edu.cn (K. Chen); jiangtr5@mail.sysu.edu.cn (T. R. Jiang); yupeng9@mail.sysu.edu.cn (P. Yu)

**Abstract:**

The unique energy band and crystal structure of the layered type-II Weyl semimetal TaIrTe$_4$ hold great promise for high-performance broadband anisotropic optoelectronic devices. Therefore, gaining an in-depth understanding of the interactions between internal microscopic particles is of vital importance. Here, we employed a two-color pump-probe system to reveal the anisotropic electron-phonon coupling (EPC) and coherent phonon dynamics in bulk TaIrTe$_4$. The carrier relaxation exhibits a four-exponential decay process, with strong dependence on polarization of probe pulse, indicating that EPC strength is closely related to the crystal axes (a/b- axes). In addition, we observed three coherent phonon modes in bulk TaIrTe$_4$: 38.5 GHz, 0.44 THz and 1.29 THz. Their oscillation amplitudes and dephasing times also showed anisotropic responses to the probe polarization. We also investigated the in-plane cross-directional thermal conductivity coefficient of TaIrTe$_4$ by beam-offset frequency-domain thermal reflection (FDTR). The thermal conductivity coefficient along the a-axis and b-axis directions are $k_a$=14.4 W/mK and $k_b$=3.8 W/mK, respectively. This represents a significant in-plane anisotropy. Our work not only reveals the key role of anisotropic EPC in controlling the thermal and optical properties of TaIrTe$_4$, but also provides insights into designing polarization-sensitive optoelectronic devices based on topological semimetals.

**Introduction**

The discovery of Weyl semimetals has ushered in a new era for condensed matter physics. These materials exhibit a range of intriguing properties, such as ultra-high mobility[1], giant magnetoresistance[2], chiral anomaly[3], axial gravitational anomaly[4] and anisotropic nonlinear optical responses[5], making the study of Weyl semimetals a hot topic. Since 2016, researchers have utilized angle resolved photoelectron spectroscopy (ARPES) to verify the existence of various new Weyl semimetals, including Ta$_3$S$_2$[6] and TaIrTe$_4$[7]. Among them, the layered crystal structure of TaIrTe$_4$ facilitates transport experiments and device development applications, while its intrinsic space symmetry breaking endows it with broad application prospects in polarization and broadband photodetectors. These superior properties of TaIrTe$_4$ have aroused great research interest among scientists. Compared with other Weyl semimetals, TaIrTe$_4$ demonstrates superior performance and greater research potential. For instance, it has been found to exhibit a nonlinear Hall effect[8] and a huge infrared light response[9] at room temperature. In 2020, F. L. Mardele et al.[10] determined the anisotropic optical conductivity of TaIrTe$_4$ and found it to be consistent with the tilted Weyl cone model at low energies. Despite numerous studies highlighting the excellent optoelectronic properties of TaIrTe$_4$ materials, most current research focuses on steady-state characteristics (such as photoconductivity, superconducting properties[11] and wide-spectrum response[12]), while the ultrafast processes of nonequilibrium photoexcited carriers (which are more relevant to optoelectronic responses and applications) remain underexplored.

Recently, Sun et al.[13] studied the infrared ultrafast spectrum of TaIrTe$_4$ and found that after ultrafast photoexcitation, the anisotropy of its optical conductivity significantly weakened (the reflectivity tended to be isotropic), this tunable property is advantageous for polarization-resolving detectors. The dynamical results show that three time-components remain unchanged when fitting different pump and probe polarizations, indicating that the carrier relaxation process tends to be isotropic. However, according to the DFT results presented in previous works,[14] there is a large dispersion of the TaIrTe$_4$ energy band in the k$_x$ (a-axis) direction of the Brillouin zone, while it is relatively small in the k$_y$ (b-axis) direction. Therefore, the effective phase space for electron-phonon scattering along the k$_x$ and k$_y$ directions should be anisotropic. This means, the cooling process of high-energy hot carriers scattered by electron-phonon interactions in different directions should be anisotropic as well. But this was not observed in their experiments, probably due to the low carrier energy they excited. Meanwhile, the intrinsic symmetry breaking of its lattice will also lead to different propagation characteristics of phonons in different directions, thereby causing anisotropic thermal conductivity. 'To the best of our current knowledge, the properties of TaIrTe$_4$ in this regard are still unclear and await further revelation.

Herein, our present work here for the first time combines ultrafast spectrum with thermal conductivity measurements to demonstrate the anisotropic EPC and thermal transport. Pump-probe reflectance spectrum reveals the carrier dynamic response of TaIrTe$_4$ under photoexcitation, which is divided into four processes: fast relaxation components ($\tau_1$, $\tau_2$): corresponding to strong EPC scattering (picosecond scale). Slow relaxation components ($\tau_3$, $\tau_4$): involving lattice heat accumulation and thermal diffusion (hundreds of picoseconds to nanoseconds scale). It was found that the carrier dynamical signals are insensitive to the pump polarization but significantly depend on the probe polarization. In addition, we observed three coherent phonons: a CAP at 38.5

GHz and two COPs at 0.44 THz and 1.29 THz, and confirmed their respective generation mechanisms. The phonon dephasing times were quantified by damped harmonic oscillator model, and they all showed strong correlation with the crystal axes. Finally, the in-plane thermal conductivity of TaIrTe$_4$ was obtained by beam-offset FDTR technique, revealing a large anisotropy ratio of 3.8. Through multi-timescale dynamics and anisotropy correlation analysis, as well as molecular dynamics calculations, new ideas for the ultrafast optical and thermal applications of topological quantum materials are provided.

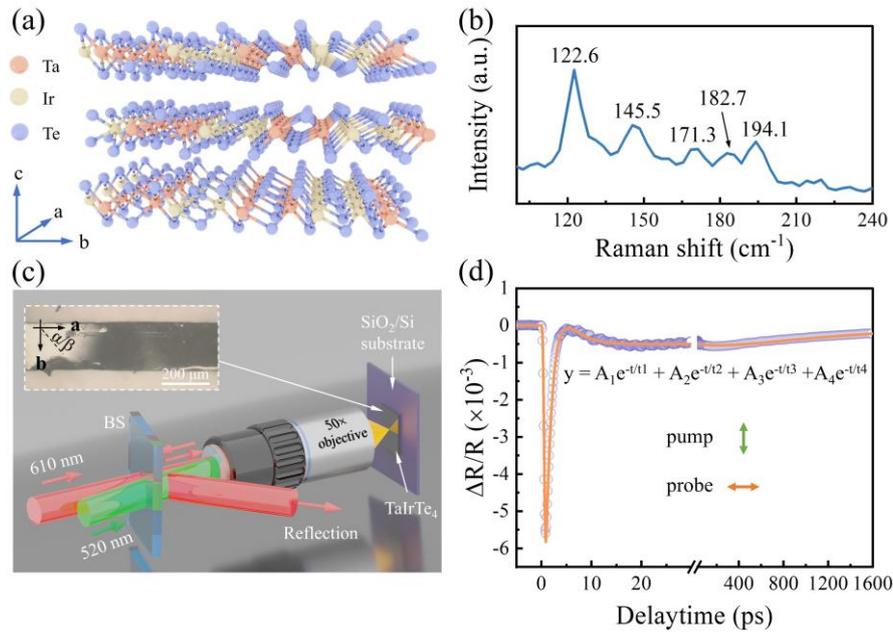

**Fig.1 Lattice structure, Raman spectroscopy, experimental scheme and typical transient reflection spectrum.** (a) Crystal structure of type-II Weyl semimetal TaIrTe$_4$. Red, yellow and blue spheres represent Ta, Ir and Te atoms, respectively. (b) The Raman spectrum of bulk TaIrTe$_4$. (c) Schematic diagram of transient differential reflection experiment. Here the pump and probe light are collinear. For easy viewing, they have been drawn separately. Inset is the optical microscope image of bulk TaIrTe$_4$. To investigate the dynamic anisotropic response of TaIrTe$_4$, a half-wave plate was used to rotate the polarization of the pump and probe. The polarization angle of 0 degrees (for pump $\alpha$ and probe $\beta$) was aligned along the crystallographic a-axis. (d) Transient differential reflection signal with parallel-polarized pump (520 nm) and probe (610 nm) geometry, the orange curve represents the fitting result. The temperature is 300 K and the pump fluences is 1.59 mJ/cm$^2$.

In the experiment, TaIrTe$_4$ single crystals were grown by a self-flux method. The details of the crystal growth process can be found in the **"Materials and Methods"** section. TaIrTe$_4$ crystallizes in an orthorhombic structure with the space group Pmn2$_1$ (No. 31)[15]. As shown in **Figure 1a**, the side view of the TaIrTe$_4$ lattice structure is presented, with red, yellow, and blue spheres representing Ta, Ir, and Te atoms, respectively. The lattice constants associated with the three crystal axes (a, b, c) are 3.770 Å, 12.421 Å, and 13.184 Å[13], respectively. The crystal structure of TaIrTe$_4$ is formed by stacking single layers in an AB order through van der Waals bonds, and can be considered as a variant of the WTe$_2$ structure (along the b direction), where Ta and Ir atoms alternately construct metal-metal chains in a zigzag arrangement[16]. The optical microscope image of bulk TaIrTe$_4$ is shown in the inset of **Figure 1c**. A representative Raman spectrum is displayed in **Figure 1b**, with Raman peaks in good agreement with previous reports[15]. Notably, the two Raman modes, A$_1$ (~145.5 cm$^{-1}$) and A$_2$ mode (134 cm$^{-1}$), are considered as signatures of spatial symmetry breaking in TaIrTe$_4$.

The optical measurement setup is illustrated in **Figure 1c** (a more detailed optical path diagram is shown in **Figure S1**). A dual-color pump-probe system (with a pump wavelength of 520 nm and a tunable probe wavelength ranging from 560 nm to 660 nm) was employed to measure the transient differential reflectance signal ($\Delta R/R_0$) of bulk TaIrTe$_4$, where R and R$_0$ represent the reflectance before and after the pump excitation, respectively. The energy of the pump light is sufficiently large to fully excite carriers from the ground state to a high-energy state, and the slightly-lower-energy probe light was used to detect the relaxation information of free carriers. The pump and probe beams were collinearly focused onto the sample surface through a 50× objective lens, with a spot diameter of 1 μm. To investigate the dynamic anisotropic response of TaIrTe$_4$, the polarization directions of the pump and probe beams were rotated using half-wave plates. The polarization 0° angle was set along the crystallographic a-axis (see the inset of **Figure 1c**).

**Figure 1d** presents a representative transient reflectance spectrum under orthogonal polarization of the pump-probe (610 nm) beams. At time zero, the bulk TaIrTe$_4$ was excited, causing electrons to be excited to higher energy states and creating a non-equilibrium carrier distribution. At this moment upon pump excitation, an immediate drop, i.e., a sharp decrease in the probe reflectance was observed. The response time was limited by the 100fs laser pulse width. This was followed by a rapid exponential decay and then a slow recovery process. We fitted the dynamic relaxation curve with a tetra-exponential decay function, which exhibited a better fit than the tri-exponential function (see **Figure S2**). This suggests that the relaxation process of photoexcited carriers involves four characteristic time-scale. After pump excitation, the photoexcited carriers first thermalize and establish a thermal equilibrium state among carriers within the laser pulse duration. Subsequently, the hot carriers cool down by transferring energy to the lattice. Similar to previous reports[13], the dynamics can be decomposed into two fast relaxation components and two slow relaxation processes: the fast relaxation component $\tau_1$ is attributed to the strong coupling scattering process between electrons and high-energy optical phonons. In this process, hot electrons rapidly transfer energy to the lattice through intense interactions with high-energy optical phonons, forming an initial cooling channel. The time constant of this scattering mechanism is typically in the picosecond range, reflecting the ultrafast energy relaxation characteristics of the electron-phonon strongly coupled system[17]. The secondary relaxation component $\tau_2$ corresponds to the scattering processes involving electrons with low-energy optical phonons and acoustic phonons. After high-energy phonon scattering, electrons continue to release energy through weak interactions with low-energy phonons. This process, due to the lower phonon energy and weaker coupling strength, results in a reduced relaxation rate and forms a time difference. Then follows a lattice temperature establishment process (via phonon-phonon scatterings) on the order of a hundred picoseconds before entering the nanosecond-scale long relaxation stage, the thermal diffusion process within the TaIrTe$_4$ system — that is, the non-local diffusion of pump-area energy into the adjacent external lattice, which is limited by the thermal conductivity rate and exhibits nanosecond-scale dynamics.

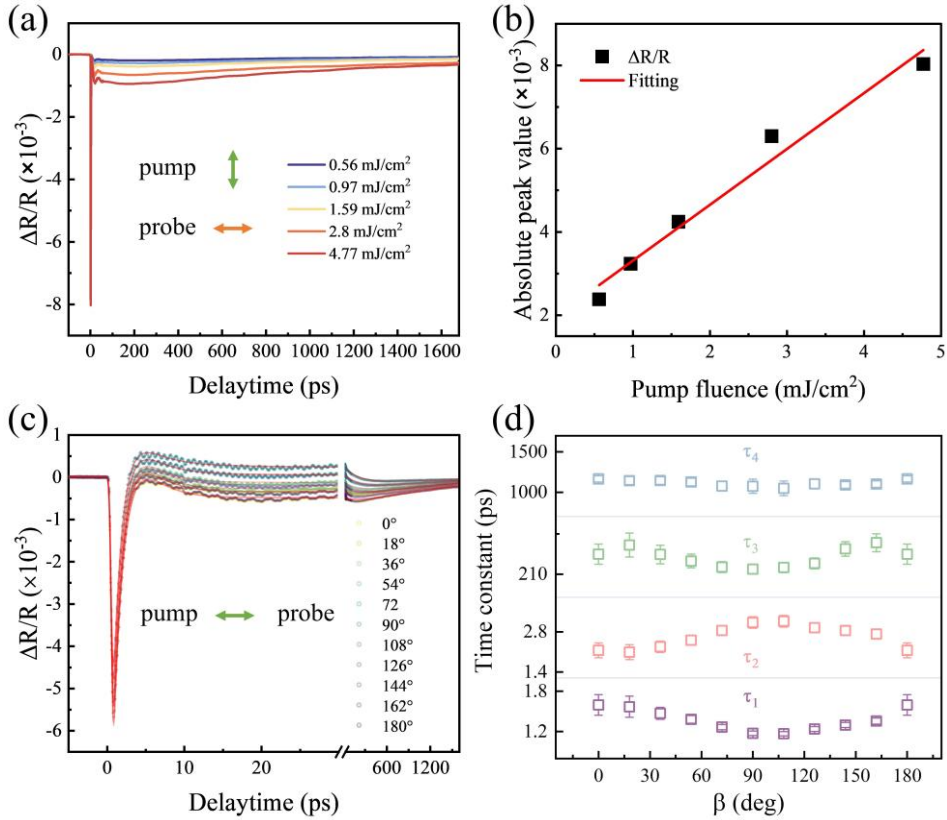

**Fig.2 Pump fluence and probe polarization dependence of transient dynamic.** (a) The dependence of transient reflectance on pump fluence with pump (α=90°) and probe (β=0°) polarization at room temperature. (b) The dependence of ΔR/R at t=0 ps on pump fluence. The experimental data can be linearly fitted by the red line. (c) Probe polarization dependence of ΔR/R with α=0° pump polarization. All curves are fitted by tetra-exponential function like fig. 1d. The pump fluence is 1.59 mJ/cm². (d) The fitted $\tau_1$, $\tau_2$, $\tau_3$ and $\tau_4$ as a function of probe-polarization angle β. The temperature is 300 K.

**Figure 2** shows the dynamic evolution under different pump fluence and probe polarization states. As shown in **Figure 2a**, under cross-polarized conditions (i.e., α = 0°, β = 90°), the ΔR/R decreases with the increase of pump fluence density. Within the studied power range, the amplitude of the negative peak at t = 0ps shows a linear relationship with the pump fluence (**Figure 2b**). This linear response indicates that system did not exhibit saturated absorption or two-photon absorption phenomena. Quantitative analysis of the decay time constants obtained from the tetra-exponential fit (see **Figure S3**, **Table S2**) shows that $\tau_1$, $\tau_2$, and $\tau_3$ increase with the increase of pump energy flux density, while $\tau_4$ exhibits a sudden drop under the maximum energy flux density, but it remains essentially constant at low energy flux densities. As although TaIrTe₄ has a large lattice heat capacity[13], the pump fluence range covered in this experiment is sufficient to cause significant lattice temperature rise (the lattice temperature calculated by the two-temperature model under different pump fluence is shown in **Table S3**). The sufficiently-high lattice temperature under the maximum energy flux density may accelerate the thermal diffusion process, as shown in a recent work for a similar material with complex structure[18].

**Figure 2c** depicts the evolution of the transient reflection spectra as a function of the probe polarization angle ($\beta$) with the pump polarization fixed along the a-axis ($\alpha = 0°$). (Results for pump polarization along the b-axis ($\alpha = 90°$) are provided in **figure S4**). Due to the symmetry breaking of the crystal structure, the ultrafast carrier dynamics are expected to exhibit anisotropic characteristics. Experimental results confirm this anisotropy: as the probe polarization rotates from 0° (a-axis) to 90° (b-axis), the $\Delta R/R$ at t = 0 ps remains negative, indicating pump-induced reduction in reflectivity along both the a- and b-axes. However, only a few picoseconds later, $\Delta R/R$ transitions to positive values near the b-axis polarization angle ($\beta \approx 90°$). Notably, a similar probe-polarization-dependent spectral evolution is observed when the pump is polarized along the b-axis, demonstrating that this phenomenon is independent of the pump polarization direction. This suggests the presence of different relaxation routes for the excited hot electrons moving in different directions. Theoretically, the $\Delta R/R$ signals reflect the changes in elements of optical conductance tensor[13]. For crystals with $C_{2v}$ symmetry like TaIrTe$_4$, while the probe light polarized along y-direction (b-axis) can only relate to the optical conductance along y-direction ($\sigma_{yy}$), the probe light polarized along x-direction (a-axis) can both sense the optical conductance along x-direction ($\sigma_{xx}$) and the off-diagnal term (c-axis, $\sigma_{zx}$). According to the Kubo Greenwood formula[19], interband $\sigma_{xx}$ and $\sigma_{yy}$ should be mainly due to the coupling between electrons of initial and final states with velocity along x- and y- direction, respectively. And $\sigma_{zx}$ is determined by the product of the $V_x$ and $V_z$ matrix elements coupling the initial and final electron states. Therefore, we can roughly derive the following conclusion: Light polarized in the y direction primarily measures electrons moving in the y direction, while light polarized in the x- direction can sense electrons moving in both the x- and z- directions. With this detection mechanism, **Fig. 2c** indicates that electrons moving in x- and y- directions relax in different ways and rates.

Spectroscopic results investigating pump polarization dependence further support this conclusion. **Figure S5a** shows normalized transient reflectivity (peak-normalized at t = 0) as a function of pump polarization angle ($\alpha$) with the probe polarization fixed at $\beta = 90°$. The normalized spectra for different pump polarization angles nearly overlap. Corresponding tetra-exponential fitting results (**Figure S5b, c**, **Table S5**) reveal extremely weak dependence of both the decay amplitudes ($A_1$–$A_4$) and time constants ($\tau_1$–$\tau_4$) on the pump polarization angle. Likewise, no significant dependence is observed with the probe polarization fixed at $\beta = 0°$, either. (**Figure S6**, **Table S6**). In contrast, the probe-polarization-dependent dynamic signals exhibit pronounced anisotropic behavior in the carrier relaxation (**Figure 2d**, **Table S7**). Specifically, the characteristic time constants $\tau_1$ and $\tau_2$, associated with electron-phonon interactions, display opposing trends as the probe polarization rotates from 0° to 90°. The time constant $\tau_2$, related to low-energy optical phonon and acoustic phonon scattering process, reaches its maximum value when the probe polarization is along the b-axis ($\beta = 90°$). This behavior coincides with the lower thermal conductivity ($k$) along the b-axis of TaIrTe$_4$ crystals (the detailed analysis will follow later) and represents an indicator that electrons moving along b-axis couple to acoustic phonons weaker than those moving along a-axis do. In summary, the pronounced anisotropy in the carrier dynamical signal of TaIrTe$_4$ makes it an ideal candidate material for developing high-speed, polarization-sensitive optoelectronic devices. Furthermore, its semi-metallic, gapless electronic structure enables broadband response capabilities, extending the operational spectrum into the mid-infrared to terahertz regions.

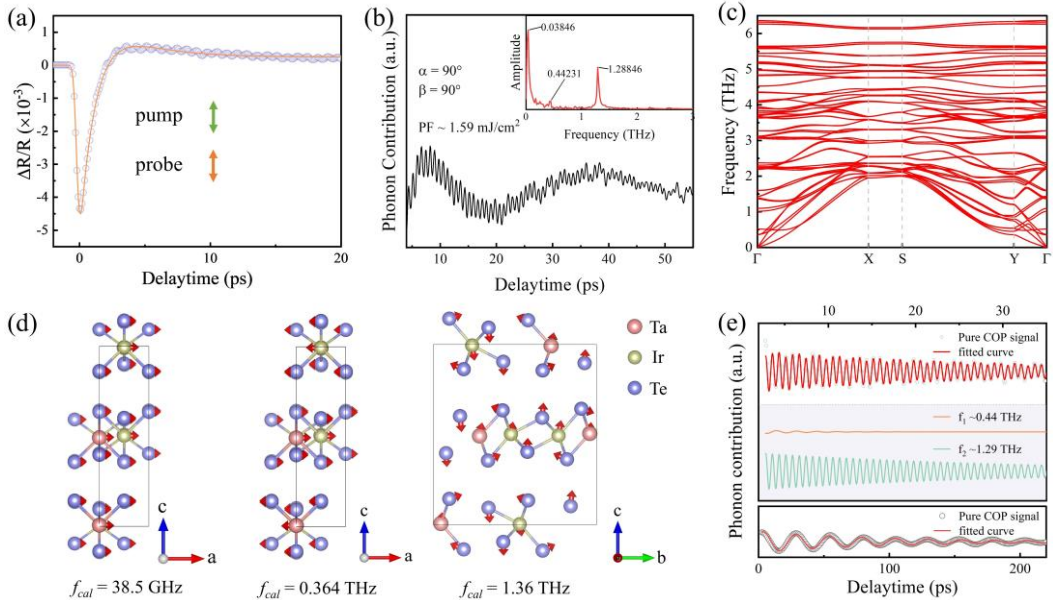

**Fig.3 Analysis of the generation mechanism of CAP and COPs.** (a) The transient differential reflection signal in the direction of the strongest coherent phonon oscillation (pump and probe are parallel-polarized along b-axis). The pump fluence (PF) is 1.59 mJ/cm$^2$, the temperature is 300 K. (b) Typical phonon contribution (black line) to the transient differential reflection signal (a), inset is its amplitude spectrum after Fourier transform. (c) Phonon dispersion spectrum of bulk TaIrTe$_4$. High-symmetric $q$-point paths: Γ (0, 0) → X (1/2, 0) → S (1/2, 1/2) → Y (0, 1/2)

→Γ (0, 0). (d) The vibration directions of three phonon frequencies moving along a-axis. Three frequencies obtained from transient reflectance spectrum, 38.5 GHz, 0.44 THz and 1.29 THz, correspond to the calculated frequencies ($f_{cal}$), 38.5 GHz, 0.364 THz, and 1.36 THz, respectively. It is evident that the 38.5 GHz CAP is transverse, the low-frequency optical branch at 0.44 THz ($f_{cal}$ = 0.364 THz) is longitudinal, and the high-frequency optical branch at 1.29 THz ($f_{cal}$ = 1.36 THz) is transverse. (e) Pure COP and CAP contribution (green and black scatter, respectively) to the transient differential reflection signal (a), two damped cosine functions are used to fit the pure COP (red line above), separating the contributions of the two optical phonons 0.44 and 1.29 THz (orange and green line, respectively), pure CAP can also be fitted with a damped cosine function (red line below).

By examining **Figure 2a**, it can be observed that in the initial stage of decay, distinct large-period oscillations (tens of picoseconds) are superimposed on the carrier dynamics signal. Moreover, under high pump fluence, some clear small-period oscillation signals (a few picoseconds) gradually emerge from the large-period oscillations, indicating the excitation of coherent phonons. However, due to their weak intensity, high-frequency oscillations could not be confirmed. Therefore, we conducted fine scans and eventually discovered that when the probe polarization is along the b-axis (β=90°), clear high-frequency coherent phonon oscillations can be observed across a broad range of prob wavelengths, as shown in **Figure S7a, c**. In contrast, at β=0°, as seen in **Figure S7b, d**, the fine structure is significantly reduced. The low-frequency oscillations, however, persist. Because probe lights with different polarizations can sense lattice vibration along different directions in different weights via the complex Raman tensor, the significant dependence of the oscillation signal on probe-polarization indicates that the pump-excited multiple coherent phonons vibrating in two orthogonal lattice directions possess distinct characteristics.

To elucidate the mechanisms behind these coherent phonons, we subtracted the carrier dynamics decay background (indicated by the red solid line in **Figure 3a**) from the transient reflectance spectra, thereby revealing the pure phonon contributions, as shown in **Figure 3b**. Utilizing Fast Fourier Transform (FFT), we calculated the primary oscillation frequencies to be approximately 38.5 GHz, 0.44 THz, and 1.29 THz (as depicted in the inset of **Figure 3b**). Regarding the ~38.5 GHz CAP oscillation (coherent lattice vibrations of acoustic modes modulate the probe light), frequencies that are of comparable magnitude have been observed in many other materials[20]. And two high-frequency oscillations can be attributed to $A_1$-COP modes within the plane, phonon dispersion spectrum of bulk TaIrTe$_4$ (**Figure 3c**) and previous report[15] has proved this point. The phonon spectrum of bulk TaIrTe$_4$ was obtained through molecular dynamics

(MD) simulations. By training a new potential function and iterating continuously, we achieved a result that matches the experimental outcomes. The comparison between the calculated frequencies and the Raman spectroscopy experimental values is shown in **Table S8**. Meanwhile, the three frequencies obtained from transient reflectance spectrum, 38.5 GHz, 0.44 THz, and 1.29 THz, can also be found in the calculation results as $f_{cal}$ = 38.5 GHz, 0.364 THz, and 1.36 THz (**Table S9**), respectively, with acceptable discrepancies. Additionally, the vibration modes of these phonons moving along the a-axis (b-axis) are depicted in **Figure 3d** (**Figure S8a**). It is evident that along the a-axis direction, the 38.5 GHz and 1.29 THz ($f_{cal}$ = 1.36 THz) phonons are transverse, while the 0.44 THz ($f_{cal}$ = 0.364 THz) phonon is longitudinal. Along the b-axis direction, the 38.5 GHz phonon is transverse, the 0.44 THz ($f_{cal}$ = 0.35 THz) phonon is transverse, and the 1.29 THz ($f_{cal}$ = 1.36 THz) phonon has both transverse and longitudinal components. The MD simulation result not only provides a clearer understanding of these three phonon modes, but also serves as strong evidence for the anisotropic phonon dynamics discussed below.

Generally, COPs are primarily generated through two mechanisms: Impulsive Stimulated Raman Scattering (ISRS)[21] and Displacive Excitation of Coherent Phonons (DECP)[22]. In ISRS, the generation of coherent phonons is first achieved through inelastic Raman scattering of photons with ground-state electrons, which requires the presence of Raman-active vibrational modes in the crystal. Secondly, ISRS excites phonons of all modes (both fully symmetric and non-symmetric), and COPs in transparent substance are generally produced by the ISRS mechanism, because ISRS does not require light-absorbing. In contrast, the DECP mechanism is a coherent phonon process based on excited-state electrons, thus necessitating strong absorption of the pump light and the presence of electron-phonon coupling in the system. Moreover, DECP only drives fully symmetric modes. The initial phase of the phonon can be used to distinguish between these two mechanisms[23]. We filtered out the contributions of the COP and CAP parts in **Figure 3b** using FFT filters to facilitate better fitting and analysis, resulting in the pure COP (indicated by green circles) and CAP (indicated by black circles) contributions shown in **Figure 3d**. The COP trace can be well fitted with two damped harmonic oscillator functions (red curve in **Figure 3d**). In our results, these two oscillations in the COP signal are both represented by cosine functions (orange and green curves in **Figure 3d**), and after fitting, the oscillation frequencies match those well obtained by FFT, both being fully symmetric $A_1$ modes. These are very typical characteristics of the DECP mechanism.

In principle, the Deformation Potential Coupling (DPC) and TE are commonly used to explain the generation of CAP oscillations in transient reflectance spectra. Since these two mechanisms operate differently, they can be described using different functions[24]: (1) DPC is described by $A_0 e^{-t/\tau_{AP}} \sin(2\pi f t - \varphi_D)$, where $\varphi_D = \arctan\left(\frac{2\pi f}{1/\tau_2 - 1/\tau_{AP}}\right)$ is the initial phase of the oscillation, $A_0$ is the amplitude at time zero, $\tau_{AP}$ is the decoherence time, $\tau_2$ is the scattering time of electrons with low-frequency optical phonons or acoustic phonons and $f$ is the oscillation frequency; (2) TE is described by $A_0 e^{-t/\tau_{AP}} \cos(2\pi f t - \varphi_T)$, where $\varphi_T = \arctan\left(\frac{1}{2\pi f \tau_{AP}} + \frac{2\pi f}{1/\tau_2 - 2/\tau_{AP}}\right)$ is the initial phase of the oscillation, $A_0$ is the amplitude at time zero, $\tau_{AP}$ is the decoherence time, and $f$ is the oscillation frequency. If the CAP oscillations can all be fitted with above sine and cosine functions, then by comparing the experimentally fitted $\varphi_{D,exp}$ ($\varphi_{T,exp}$) with the theoretical values $\varphi_D$ ($\varphi_T$), the mechanism can be determined. When $f = 38.5\ GHz$, $\tau_{AP} = 97\ ps$, $\tau_{AP} = 3.5\ ps$, we obtained $\varphi_{D,exp} = -39°$ ($\varphi_D = 42°$) and $\varphi_{T,exp} = 50°$ ($\varphi_T = 44°$). The value of $\varphi_T$ is closer to the experimental value $\varphi_{T,exp}$, so the Thermoelastic Effect is the predominant mechanism for generating CAP in bulk TaIrTe$_4$.

The strong anisotropic electroacoustic coupling exhibited in the carrier dynamical signal suggests that the phonon dynamics should also possess anisotropic properties. Therefore, the polarization dependence of the coherent phonon spectrum was investigated. We first subtract the carrier part from the pump polarization-dependent dynamics (**Figures S5a** and **S6a**) as shown in **Figures S9a** and **S9b**. The results obtained by FFT (**Figures S9c-f**) demonstrate that the frequency and amplitude of these coherent phonons are insensitive to the pump polarization, which further confirms that the generation mechanism of COP is not ISRS, as ISRS requires sensitivity to pump polarization[25]. For DECP, although the initial excitation by the pump light establishes an anisotropic carrier distribution in k-space relative to the pump polarization, the momentum of the photoexcited carriers is rapidly randomized through carrier scattering, and the influence of the pump polarization completely disappears within tens of femtoseconds[26]. Therefore, unlike ISRS, DECP, which is controlled by the photoexcited carrier ensemble, should have lower sensitivity to the pump pulse polarization in the excitation of coherent phonon oscillations, consistent with the experimentally measured pump polarization-dependent coherent phonon amplitudes (**Figures S9e, f**). Furthermore, the fitting of the pure COP spectrum also indicates that the decoherence time is almost isotropic (**Figures S9g, h**).

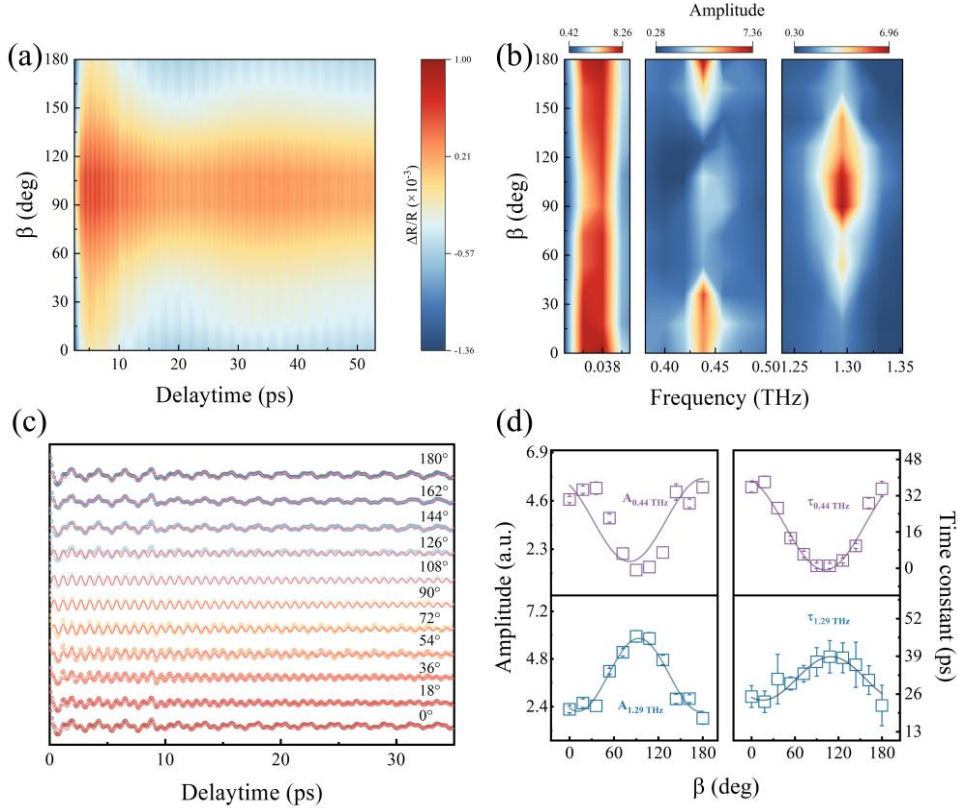

**Fig.4 Anisotropic ultrafast phonon dynamics.** (a) 2D color plot of time-resolved TR traces of bulk TaIrTe$_4$ under different probe polarizations at pump fixed at α=0°. (b) 2D color plot of the amplitudes of each mode (38.5 GHz, 0.44 and 1.29 THz) under different probe polarizations. (c) COP oscillation spectrum (scatters) and fitting curves (red lines) under 0-180° probe-polarization. (d) The dependence of time constants and amplitude on probe-polarization, a cosine function can fit them well.

The significant differences observed in the orthogonal direction of probe light previously indicate that there is indeed a large in-plane anisotropy in the coherent phonons. Therefore, we fixed the pump polarization along a-axis (α=0°) and performed fine scans by changing the probe polarization at intervals of 18°, as shown in **Figure 4a**. It can be seen that the fine oscillation fringes are clearly visible near β=90° in the figure, while near β=0° or 180°, these fine light and dark fringes are much more blurred, obviously affected by the probe polarization. After removing the contribution of carrier dynamics, the FFT results of the coherent phonon spectrum show that the three coherent phonon oscillation frequencies of ~38.5 GHz, ~0.44 THz, and ~1.29 THz are not affected by the probe polarization (**Figure S10a**), while the oscillation amplitude has a strong anisotropy with the probe polarization (**Figure 4b**). It can be well fitted by a cosine function, as shown in **Figure S10b**, and it can be anticipated that when extended to a range of 360°, it will exhibit double symmetry. As the probe polarization changes, the amplitude of the high-frequency COP reaches its maximum along b-axis, while the amplitude of the CAP and low-frequency COP approach the minimum, with the low-frequency COP almost disappearing near 90°. This phenomenon is predictable. As shown in **Figures 3d** and **S8a**, low-frequency COP vibrates along the a-axis in both directions, thus its amplitude is maximized when the probe-polarization is aligned with a-axis. In contrast, the high-frequency COP has negligible oscillation components along the a-axis but possesses oscillation components along the b-axis, resulting in a maximum amplitude when probe-polarization along b-axis. However, for CAP, the detected vibration conditions are the result of the combined action of several acoustic branches (figure 3d and figure S8b), so the results show a relatively weak anisotropic response, with the amplitude along the a-axis being relatively larger. This indicates that a-axis and b-axis are each favorable for different coherent phonon vibrations. This is due to the symmetry breaking of the TaIrTe$_4$ itself, the lack of inversion symmetry in the orthorhombic system, leading to the intrinsic asymmetry of phonon vibration modes along different crystal axes (a/b-axis), as well as the directional selectivity of electron-phonon coupling. **Figure 4c** shows the extracted pure COP oscillation spectrum dependent on the probe polarization, where both modes exist near 0° and 90°, but near 90° it becomes a single mode. The contributions of the two COPs were separated using two damped harmonic oscillator functions, as shown in **Figure 4d**, and the anisotropic phonon oscillation amplitude is consistent with the results obtained by FFT.

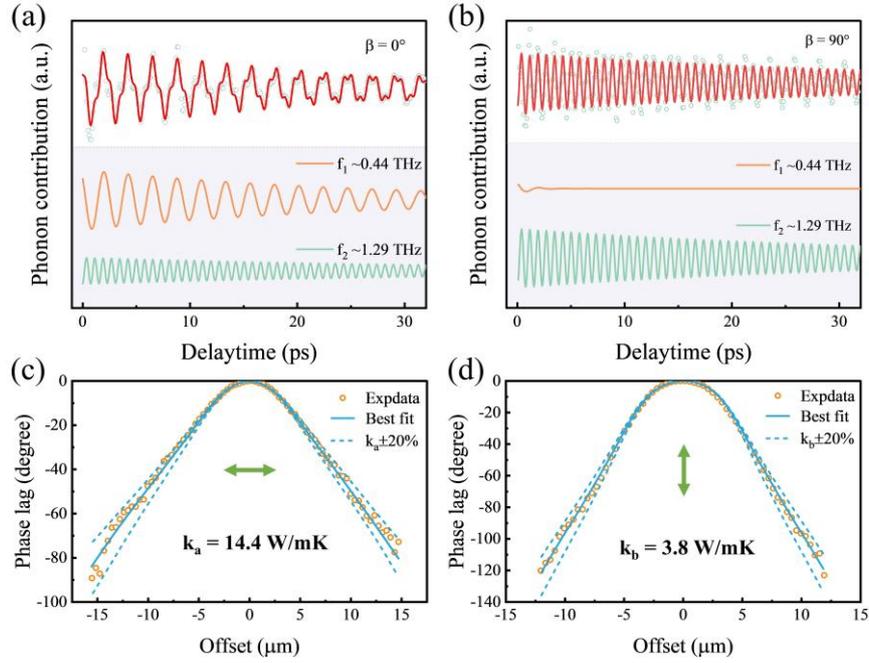

**Fig.5 Anisotropic thermal coefficient.** Decomposition of COP contribution when the probe polarization is respectively along the a-axis (a) and b-axis (b) directions, pump polarization is fixed along the a-axis. (c, d) Experimental data (red scatter) and model fitting (blue line) for the bulk TaIrTe$_4$ with phase lag signals. The measurements are conducted using a modulation frequency of 51 kHz at room temperature. (c) is measured along the b-axis, (d) is measured along the a-axis.

In addition, the decoherence times (**Figure 4d**) of COPs also exhibit opposite anisotropic characteristics. There is a close relationship between the phonon decoherence time and the thermal conductivity. In heat conduction, the coherence of phonons affects their propagation efficiency, thereby influencing the thermal conductivity of the material. According to the Green-Kubo method, the thermal conductivity $\kappa$ can be calculated through the autocorrelation function of the heat flux[27]. When considering the phonon decoherence time, the expression for thermal conductivity can be written as: $\kappa = \frac{1}{3}\sum_\lambda C_\lambda v_\lambda^2 \int Cor_\lambda(t)dt$ where $C_\lambda$ is the mode specific heat, $v_\lambda$ is the group velocity of phonons, and $Cor_\lambda(t)$ is the phonon decay function. When considering the phonon decoherence time, the decay function can be expressed as: $Cor_\lambda(t) = e^{-t/\tau_P} e^{-4ln2 t^2/\tau_c^2}$ where $\tau_P$ is the phonon lifetime and $\tau_c$ is the phonon decoherence time. Regarding the same mode, as the phonon decoherence time $\tau_c$ increases, the coherence of phonons is enhanced, leading to an effective contribution to thermal conductivity. As shown in **Figures 5a and 5b**, the decoherence curves of COPs at $\beta=0°$ and 90° are extracted, respectively. It can be seen that compared to the case of probe polarization at 0°, the 0.44 THz coherent phonon decoheres rapidly at 90° oscillation, while 1.29 THz phonon decoheres faster at 0° oscillation. Moreover, the 0.44 THz COP exhibits a greater difference in the two orthogonal in-plane directions.

The significant difference in dephasing time among the measured coherent phonons with different vibrational directions, suggests that the thermal conductivity of this material should also possess anisotropic nature. Hence, we characterized the in-plane thermal conductivity through FDTR[28]. For spatially resolved thermal analysis, a beam-offset FDTR configuration was employed, in which the pump beam was systematically scanned across the sample surface relative to the fixed probe beam, while the thermore-flectance signal was recorded as a function of the pump-probe bias distance at a fixed modulation frequency of 51 kHz, as shown in **Figures 5a and 5b**. Taking the volumetric heat capacity $C=1.46\times10^6$ J/(m³·K) of TaIrTe$_4$ from the literature[13], the in-plane thermal conductivities of TaIrTe$_4$ along the orthogonal direction were obtained by fitting the phase signal in the figure, with $k_a$=14.4 W/mK along the a-axis and $k_b$=3.8 W/mK along the b-axis.

In addition to the symmetry breaking of the TaIrTe$_4$ crystal structure itself, we found that the large thermal transport anisotropy may also be related to the COP we detected. The slow decay of the 0.44 THz COP along the a-axis indicates persisted coherence and minimal scattering losses in this direction, leading to an increased mean free path for this vibrational direction. Conversely, along the b-axis, the phonon decoheres rapidly, indicating electron-phonon and/or phonon-phonon scatterings are more frequent and stronger for phonons vibrating along b-axis. Even though phonon vibration and transport directions are independent, we still conjecture that the strong anisotropy in thermal transport might have some links to the strong anisotropy in phonon decoherence. It's worth noting that such strong anisotropic thermal conductivity in TaIrTe$_4$ has not been reported, and the detailed mechanism for anisotropic thermal transport is under investigation and will be discussed elsewhere. This finding of largely different in-plane thermal conductivities has important implications for designing devices such as directional thermal management, thermoelectric applications, and thermal sensors, which can significantly enhance the performance and efficiency of related technologies and equipment.

In conclusion, this study has observed strong anisotropic relaxation phenomena of carriers in the type-II Weyl semimetal TaIrTe$_4$ by all-optical pump-probe and beam-biased FDTR techniques. We have also firstly observed the linearly polarized excitation of CAP and A$_1$-COP in TaIrTe$_4$. Moreover, the observation of a large anisotropic thermal conductivity in TaIrTe$_4$, with an anisotropy ratio as high as 3.8, is also reported for the first time. In this study, we have revealed that the generation mechanisms of CAP and COP are TE and DECP, respectively. The strength of electron-phonon coupling and the anisotropic oscillation amplitudes and decoherence times of CAP and COP are all closely related to the probed crystal axes. Specifically, the coherent phonon modes exhibit remarkably different decoherence time along different axes, potentially linking the dynamical vibrational properties to the anisotropic thermal conductivity. This work not only provides a deeper understanding of the ultrafast carrier and phonon dynamics in type-II Weyl semimetals, but also offers important insights for the design of polarization-sensitive optoelectronic and thermal management devices based on topological semimetals.

## MATERIALS AND METHODS

**Sample preparation for bulk TaIrTe$_4$**

TaIrTe$_4$ single crystals were synthesized via the self-flux method using high-purity elemental Ta powder (99.99%), Ir powder (99.999%), and Te lump (99.999%) in an atomic ratio of 1:1:12. The stoichiometric mixture was loaded into a quartz tube and flame-sealed under a high vacuum of $10^{-6}$ Torr. The tube was then heated to 1323 K and held at this temperature for 7 days, followed by slow cooling to room temperature over 15 days to ensure a complete reaction. Subsequently, the samples were transferred to another evacuated quartz tube lined with quartz wool under a vacuum of $10^{-6}$ Torr and centrifuged at 873 K to separate the TaIrTe$_4$ crystals from the excess Te. Finally, bar-shaped TaIrTe$_4$ single crystals with a metallic luster were obtained.

**Time-resolved transient reflection measurements and Raman spectroscopy**

A femtosecond Ti: sapphire laser source (Coherent, Chameleon Discovery, 80 MHz repetition rate, 100 fs pulse width) was employed. This laser system features two ports: a fixed port and a tunable port. The fixed port generates linearly polarized laser pulses at a wavelength of 1040 nm, which are frequency-doubled to 520 nm using a BBO crystal to serve as the pump pulse. The tunable port produces wavelengths ranging from 1120 to 1320 nm, which, after passing through a frequency doubler, yields tunable probe light with wavelengths between 560 and 660 nm. The linear polarization directions of the pump and probe pulse are independently adjusted by half-wave plates. A non-polarizing beam splitter is used to combine the pump and probe beams, which are then focused onto bulk $TaIrTe_4$ using a Mitutoyo, M Plan NIR 50× objective ($N_A$= 0.42), resulting in a spot size of approximately 1 μm. The reflected probe light signal is collected by a sensitive photodetector, with the collinear pump pulse filtered out by a filter. A lock-in amplifier, modulated by a mechanical chopper at a reference frequency of 1333 Hz, is employed to read the transient reflectance signal of the probe pulse. For imaging of the sample and laser spot, an additional beam splitter directs the reflected light into a CMOS sensor.

The Raman spectra were measured by using a WITec alpha-300R confocal microscope equipped with a spectrometer (UHTS 300 SMFC VIS). The 532 nm excitation light was focused by an ×100 objective (Zeiss, 0.9 NA) into a spot diameter of about 0.8 μm to excite the samples.

**Beam-biased FDTR**

Beam-biased FDTR[28] is a thermal property measurement technique based on the pump-probe principle. It employs two laser beams (pump and probe) for testing, where the pump beam is used to heat the sample, and the probe beam is used to detect the thermal reflection signal. During the measurement, a fixed modulation frequency (51 kHz, chosen based on the material's thermal conductivity) is used to ensure that the thermal diffusion length is more than three times the spot size. By precisely controlling the offset distance between the two beams ($x_c$), the phase difference ($\Delta\phi$) corresponding to each $x_c$ is recorded. The in-plane thermal conductivity (k) is then obtained by fitting $\Delta\phi$.

**Molecular dynamics simulation for calculating phonon spectrum and Phonon vibration mode**

In the training process of the machine learning interatomic potential, we employed density functional theory (DFT) to generate the dataset used for constructing the potential. A large number of single-point energy calculations were performed using Vienna ab initio simulation package (VASP),[29] where the projector augmented-wave (PAW) method,[30] as implemented in VASP, was used to describe the electron-ion interactions. For the electronic self-consistent loop, an energy convergence threshold of $1 \times 10^{-7}$ eV was set, and a plane-wave energy cutoff of 600 eV was adopted. The Brillouin zone was sampled using a Γ-centered k-point mesh with a spacing of 0.15 Å$^{-1}$. Once the training dataset was prepared, we used the NEP executable provided in the GPUMD package[31] to train the neural network potential (NEP) model. The resulting machine-learned potential more accurately captures the interatomic potential energy landscape compared to traditional empirical potentials. With the trained potential, we conducted molecular dynamics simulations using LAMMPS,[32] and combined the results with Phonopy[33] to compute the phonon spectrum, as well as the phonon vibrational modes and directions.[34]

**Acknowledgements**

We acknowledge the support from National Key Research and Development Program of China No. 2023YFB4603801; National Natural Science Foundation of China No. 52176173, No. 21FAA02809 and No. 12304386; Guangdong Innovative and Entrepreneurial Research Team Program No. 2021ZT09L227; GuangDong Basic and

## Author Contributions

Z. Z. and T. J. conceived the experiment. Z. Z. and T. J. conducted ultra-fast dynamic measurements. Z. Z. carried out optical measurements. H. Y. prepared and transferred the TaIrTe$_4$ onto the substrate. J. L. conducted thermal conductivity measurements. J. W. performed theoretical calculations. K. C., Z. Z., T. J., J. W., and T. L. analyzed the data, and all the authors discussed the results. Z. Z. wrote the manuscript with contributions from all authors. K. C., T. J., and P. Y. supervised the project.

## Conflict of interest

The authors declare that they have no conflict of interest.

## Supplementary Materials

The Supporting Information is available free of charge.

## Data Availability

The data supporting the findings of this study are available from K. C. or T. J upon reasonable request.